\documentclass[reprint,prd,twocolumn,superscriptaddress]{revtex4-1}

\usepackage{color}
\usepackage{amsmath}
\usepackage{graphicx}

\allowdisplaybreaks

\begin{document}
\hfill TTP20-020,  P3H-20-018
\title{The Kinetic Heavy Quark Mass to Three Loops}
\author{Matteo Fael}
\email{matteo.fael@kit.edu}
\author{Kay Sch\"onwald}
\email{kay.schoenwald@kit.edu}
\author{Matthias Steinhauser}
\email{matthias.steinhauser@kit.edu}
\affiliation{Institut f\"ur Theoretische Teilchenphysik,
  Karlsruhe Institute of Technology (KIT), 76128 Karlsruhe, Germany}
\begin{abstract}
  We compute three-loop corrections to the relation between the heavy
  quark masses defined in the pole and kinetic schemes. Using known
  relations between the pole and $\overline{\rm MS}$ quark masses we can
  establish precise relations between the kinetic and $\overline{\rm MS}$
  charm and bottom masses. As compared to two loops, the precision is improved
  by a factor two to three.  Our results constitute important ingredients for
  the precise determination of the Cabibbo-Kobayashi-Maskawa matrix
  element $|V_{cb}|$ at Belle~II.
\end{abstract}
\maketitle


\bigskip {\bf Introduction.}  Among the main aims of the Belle~II experiment
at the SuperKEKB accelerator at KEK (Tsukuba) is the precise measurement of
various matrix elements in the Cabibbo-Kobayashi-Maskawa (CKM) mixing matrix.
These are crucial ingredients for our understanding of charge-parity (CP)
violation and indispensable input for precision tests of the Standard Model
(SM) of particle physics.  In this context the determination of $|V_{cb}|$, the
CKM matrix element entering in $b \to c $ transitions, at the 1\% level is of
particular interest; at present its relative error of about
2\%~\cite{Amhis:2019ckw} constitutes an important source of uncertainty in
the predictions for $K \to \pi \nu \bar \nu$~\cite{Buras:1997fb,Brod:2010hi},
$B_s \to \mu^+ \mu^-$~\cite{Bobeth:2013uxa} and
$\varepsilon_K$~\cite{Ligeti:2016qpi}, the parameter which quantifies CP
violation in kaon mixing.  All such processes set strong constraints on new
physics with a generic flavour and CP structure.

At present, the values of $|V_{cb}|$ from inclusive $b \to c \ell \nu$ decays
are obtained from global fits of
$|V_{cb}|$, the bottom and charm masses ($m_{b,c}$) and the relevant
non-perturbative parameters in the heavy quark expansion.  The most recent
determination is $|V_{cb}| = (42.19 \pm 0.78) \times
10^{-3}$~\cite{Gambino:2013rza,Alberti:2014yda,Gambino:2016jkc,Amhis:2019ckw},
where the precision is limited by perturbative and power correction uncertainties.

In analyses of $B \to X_c \ell \nu$ decays, it is mandatory to use a so-called
``threshold'' mass, designed such that the perturbative QCD corrections to the
decay rate are well-behaved.  So far, for the analyses either the kinetic mass
($m^{\rm kin}$)~\cite{Bigi:1996si} or the $1S$
mass~\cite{Hoang:1999zc,Hoang:1998hm,Hoang:1998ng,Bauer:2004ve} have been chosen.  Both
schemes are well suited for $B \to X_c \ell \nu$, since they allow for
renormalization scales $\mu \le m_b$.  The relation between the $1S$ and
$\overline{\rm MS}$ quark mass ($\overline{m}$) has been computed up to
next-to-next-to-next-to-leading order in
Refs.~\cite{Marquard:2015qpa,Marquard:2016dcn}.  For the
$m^{\rm kin}$--$\overline{m}$ relation two-loop corrections and the
three-loop terms with two closed massless fermion loops (often referred to
as large-$\beta_0$ terms) have been computed in
Ref.~\cite{Czarnecki:1997sz}.

The rate and the moments of $B \to X_c \ell \nu$ strongly depend on the mass
definition of the heavy quark, the choice of which is closely intertwined with the
size of the QCD corrections.  Perturbative calculations using the
on-shell mass scheme are affected by the renormalon ambiguity, which
manifests itself through bad behaviour of the perturbative
series~\cite{Beneke:1994sw,Bigi:1994em}.
However, QCD corrections to the semi-leptonic rates exhibit a bad convergence also
in the $\overline{\rm MS}$ scheme~\cite{Bigi:1996si,Melnikov:2000qh}.  In
fact, large $(n \alpha_s)^k$ terms, with $n=5$, arise from the
$m^{\rm OS}$--$\overline{m}$ conversion of the overall factor
$\Gamma \simeq G_F^2 m_b^5 |V_{cb}|^2 /(192 \pi^3)$.

The kinetic scheme was introduced in~\cite{Bigi:1996si} to resum such
$n$-enhanced terms via a suitable short-distance definition. It relies on the
Small Velocity QCD sum rules~\cite{Bigi:1994ga}, which hold in the zero-recoil
limit, i.e.\ for hadronic final state velocities $|\vec v\,|\ll 1$ 
in the rest frame of the decaying particle and $m_b \sim m_c$.

Note, that the semi-leptonic $B$ decays alone precisely determine only a linear
combination of the heavy quark masses, approximately given by
$m_b - 0.8 m_c$~\cite{Gambino:2013rza}. Thus, in order to break the degeneracy
one must include in the fit external constraints for the bottom and the charm
masses, which are usually given in the $\overline{\rm MS}$ scheme.  Until now
the scheme-conversion uncertainty from $\overline{m}_b(\overline{m}_b)$
to $m_b^{\rm kin}$(1GeV) dominates the uncertainty of the
$\overline{\rm MS}$ bottom quark mass~\cite{Gambino:2011cq}.
The global fits in~\cite{Gambino:2013rza,Alberti:2014yda}
employed only $\overline{m}_c$ as external input, as the gain in accuracy with
the further inclusion of $\overline{m}_b$ would have been limited by scheme conversion~\cite{Gambino:2013rza}.

In this Letter we will present the complete three-loop corrections to the
$\overline{m}$--$m^{\rm kin}$ relation, which lead to a significant improvement
of the uncertainties in the mass conversion.  Our results constitute a
fundamental ingredient for future inclusion of ${\cal O}(\alpha_s^3)$ corrections in
semi-leptonic rates and spectral moments. Thus it is one of the major steps
towards the reduction of the theoretical uncertainties affecting the $|V_{cb}|$
determination from inclusive decays at the 1\% level or even below.


\bigskip
{\bf Kinetic mass definition.}
In Ref.~\cite{Bigi:1996si} (see also Ref.~\cite{Czarnecki:1997sz}) the
kinetic mass has been defined via its relation to the pole mass $m^{\rm OS}$
through
\begin{eqnarray}
  m^{\rm kin}(\mu) 
  &=& m^{\rm OS} 
      - \overline{\Lambda}(\mu)|_{\rm pert}
      - \frac{\mu_\pi^2(\mu)|_{\rm pert}}{2 m^{\rm kin}}
      + \ldots
  \,,
  \label{eq::mkin}
\end{eqnarray}
where the ellipses stand for contributions from higher dimensional operators.
The scale $\mu$, the so-called Wilsonian cut-off, is part of the definition of
$m^{\rm kin}$ and takes the role of a normalization point for the kinetic
mass. In practice it is of the order of 1~GeV.

\begin{figure}[t]
  \begin{tabular}{cc}
  \includegraphics[width=0.18\textwidth]{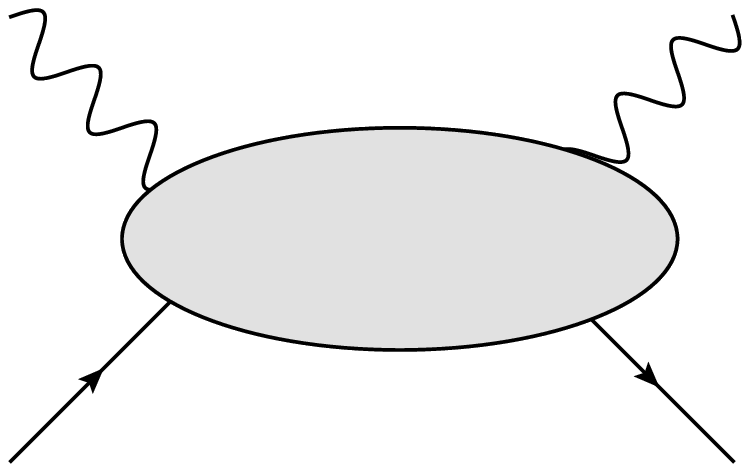} & \hspace*{2em}
  \includegraphics[width=0.18\textwidth]{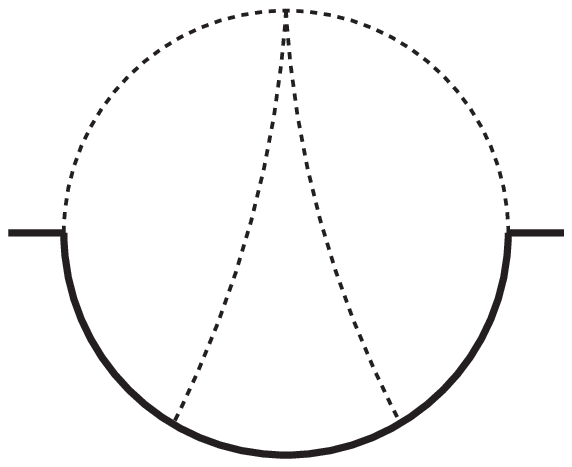}
    \\
    (a) & (b)
  \end{tabular}
  \caption{\label{fig::diag}(a) Schematic Feynman diagram representing the
    scattering of an external current, represented by wavy lines and a heavy
    quark (solid line). The blob represents one-particle irreducible quantum
    corrections, which we consider up to three-loop order.
    (b) The most complicated master integral. Dotted lines represent massless
    relativistic propagators and solid lines stand for eikonal propagators
    with mass $y$.}
\end{figure}

The quantities $\overline{\Lambda}(\mu)|_{\rm pert}$ and
$\mu_\pi^2(\mu)|_{\rm pert}$ in Eq.~(\ref{eq::mkin}) correspond to the
heavy meson's binding energy and the residual kinetic energy parameters, respectively. They are defined within
perturbation theory and are obtained from the forward scattering amplitude of
an external current $J$ and the heavy quark $Q$ [cf. Fig.~\ref{fig::diag}(a)]
\begin{eqnarray}
  T(q_0,\vec{q}\,) &=& \frac{i}{2m} \int {\rm d}^4{x \,} e^{-iqx}\langle Q|T J(x)J^\dagger(0)|Q\rangle
  \,,
  \label{eq::Tkin}
\end{eqnarray}
where for later convenience we have separated the energy and three-momentum
components of the external momentum $q$.  We furthermore denote the external
momentum of the heavy quark by $p$ with $p^2=m^2$, and we introduce
$s=(p+q)^2$.  We assume that the current $J$ does not change the flavour of
the heavy quark with mass $m$.  For $\overline{\Lambda}(\mu)|_{\rm pert}$ and
$\mu_\pi^2(\mu)|_{\rm pert}$ one has in the rest frame of the heavy
  quark~\cite{Bigi:1996si,Czarnecki:1997sz}
\begin{eqnarray}
  \overline{\Lambda}(\mu)|_{\rm pert}
  &=& 
      \lim_{\vec{v}\to0}\lim_{m\to\infty} 
      \frac{2}{\vec{v}\,^2} 
      \frac{\int_0^\mu \omega W(\omega,\vec{v}\,) {\rm d}\omega}
      {\int_0^\mu W(\omega,\vec{v}\,) {\rm d}\omega}
      \,,\nonumber\\
  \mu_\pi^2(\mu)|_{\rm pert}
  &=& 
      \lim_{\vec{v}\to0}\lim_{m\to\infty} 
      \frac{3}{\vec{v}\,^2} 
      \frac{\int_0^\mu \omega^2 W(\omega,\vec{v}\,) {\rm d}\omega}
      {\int_0^\mu W(\omega,\vec{v}\,) {\rm d}\omega}
      \,,
      \label{eq::Lam_mupi}
\end{eqnarray}
where the structure function $W$ is given by the discontinuity of $T$,
$W = 2 \mbox{Im} \left[ T(q_0,\vec{q}\,) \right]$. In Eq.~(\ref{eq::Lam_mupi})
we have $\omega=q_0-q_0^{\rm
  min}$, $\vec{v}=\vec{q}/m$ and $q_0^{\rm min}= \sqrt{\vec{q}\,^2+m^2} - m =
mv^2/2 + {\cal O}(v^4)$.
Note that $W$ is zero for $q_0<q_0^{\rm min}$.

In order to compute corrections of ${\cal O} (\alpha_s^3)$ to
Eq.~(\ref{eq::mkin}) one has to consider three-loop corrections to the
imaginary part of $T(q_0,\vec{q}\,)$ in Eq.~(\ref{eq::Tkin}). This requires
the evaluation of real and virtual corrections to the scattering process
shown schematically in Fig.~\ref{fig::diag}(a).

More details on the derivation of Eq.~(\ref{eq::Lam_mupi}) are provided in
Ref.~\cite{Fael_etal}.


\bigskip
{\bf Calculation.}
From Eqs.~(\ref{eq::mkin}) and~(\ref{eq::Lam_mupi}) we learn that the relation
between the kinetic and pole mass is obtained from the imaginary part of the
structure function $W(\omega, \vec{v}\,)$ in the limit $\vec{v}\to 0$. It is thus
suggestive to apply the threshold
expansion~\cite{Beneke:1997zp,Smirnov:2012gma}, which in our situation reduces
to two momentum regions: the loop momenta can be either hard (h) and scale as the
quark mass $m$, or ultra-soft (u) and scale as $y/m$ where $y=m^2-s$ measures the
distance to the threshold. Note that in our case we have $y<0$.  When
expanding the denominators one has to assume that both $p$ and $q$ scale as $m$.

We generate the four-point Feynman amplitudes with {\tt
  qgraf}~\cite{Nogueira:1991ex} and translate the output to {\tt
  FORM}~\cite{Ruijl:2017dtg} notation. We make sure that the external momenta
$p$ and $q$ are routed through the heavy quark line.
Afterwards we expand all loop momenta according to the
rules of asymptotic expansion which leads to a decomposition of each
integral into regions in which the individual loop momenta either scale as hard
or ultra-soft. At one-loop order there are only two regions. At two loops we
have the regions (uu), (uh) and (hh), and at three loops we have (uuu), (uuh),
(uhh) and (hhh). For each diagram we have cross-checked the scaling of the
loop momenta using the
program {\tt asy}~\cite{Pak:2010pt}. Note that the contributions where all
loop momenta are hard can be
discarded since there are no imaginary parts. The mixed regions are expected
to cancel after renormalization and decoupling of the heavy quark from the
running of the strong coupling constant. Nevertheless we perform an explicit
calculation of the (uh), (uuh) and (uhh) regions and use the cancellation as
cross check.  The physical result for the quark mass relation is solely
provided by the purely ultra-soft contributions.

The starting point of our calculation are four-point functions. However, after
the various expansions we obtain two-point functions with external momentum
$p$. As a consequence denominators become linearly dependent and a partial
fraction decomposition is needed in order to generate linear independent sets
of propagators. They serve as input for {\tt FIRE}~\cite{Smirnov:2019qkx} and
{\tt LiteRed}~\cite{Lee:2012cn} which are used for the reduction
to master integrals.

After partial fraction decomposition we end up with 1, 2 and 14 pure
ultra-soft integral families at one-, two- and three-loop order, respectively.  The
three-loop families have eight propagators and four irreducible
numerators, three of which contain scalar products of the loop
momenta and the external momentum $q$ and have been introduced to avoid an
expensive tensor reduction.

After reduction to master integrals and their subsequent minimization
across all families the amplitude can be expressed in terms of 
1, 3 and 20 ultra-soft master integrals at one-, two- and three-loop order,
respectively. At one and two loops all of them can be expressed in terms of
$\Gamma$ functions.  This is also the case for 11 of the three-loop
  master integrals.  For 8 of the remaining integrals we obtain
analytic results for the $\epsilon$ expansion with the help of
Mellin-Barnes~\cite{Smirnov:2012gma} representations.  In these cases the
residues obtained after closing the integration contour can be summed
analytically with the packages
\texttt{Sigma}~\cite{Schneider:2007},
\texttt{EvaluateMultiSums}~\cite{Ablinger:2010pb}
together with
\texttt{HarmonicSums}~\cite{HarmonicSums};
additionally we obtain high-precision numerical results and use the
PSLQ~\cite{PSLQ} algorithm to reconstruct the analytic expression. We have
only encountered one integral where a different strategy was necessary. It is
shown in graphical form in Fig.~\ref{fig::diag}(b).  For this integral we have
introduced a different mass scale, $x$, in the  bottom-middle
propagator. In case this mass is zero ($x=0$), the integral can be computed
analytically. Thus, it is suggestive to establish differential
equations~\cite{Kotikov:1990kg,Gehrmann:1999as,Henn:2013pwa}, apply boundary
conditions at $x=0$, and evaluate the solution for $x=1$, which provides the
desired integral. We will provide more details on the computation of the
master integrals in Ref.~\cite{Fael_etal}.

Let us mention that we have performed our calculation for a general  gauge
parameter $\xi$. We expand the amplitude up to linear order in $\xi$ and check
that $\xi$ cancels after adding the quark mass counterterms.  Furthermore, for
the external current $J$ we use both a vector ($J=\bar{Q}\gamma_\mu Q$) and a
scalar ($J=\bar{Q}Q$) current and check that the final result for the relation
between the pole and kinetic mass is the same.  However, the intermediate
expressions are different. This concerns, e.g., the renormalization of the
current itself. Whereas the vector current has  a vanishing anomalous dimension an
explicit renormalization constant is needed for the scalar
current. Furthermore, in the case of the vector current there is no
contribution from the virtual corrections contained in the denominator of
Eq.~(\ref{eq::Lam_mupi}) since in the static limit the Dirac form factor
vanishes and the Pauli form factor is suppressed by $q^2$.  On the other hand,
in the scalar case there is a contribution from the finite static form factor.


\bigskip
{\bf Results.}
The main result of our calculation is the relation between the kinetic and the
pole mass, which up to order $\alpha_s^3$ is given by
\begin{widetext}
\begin{eqnarray}
  \frac{m^{\text{kin}}}{m^{\text{OS}}} &=& 
    1
    - \frac{\alpha_s^{(n_l)}}{\pi} C_F
    \biggl(
      \frac{4}{3} \frac{\mu}{m^{\text{OS}}} 
                       + \frac{1}{2} \frac{\mu^2}{\left({m^{\text{OS}}}\right)^2}
    \biggr)
    + \left( \frac{\alpha_s^{(n_l)}}{\pi}\right)^2 C_F
    \Biggl\{
      \frac{\mu}{{m^{\text{OS}}}}
      \Biggl[
        C_A 
        \biggl(
          - \frac{215}{27}
          + \frac{2\pi^2}{9} 
          + \frac{22}{9} l_\mu
        \biggr)
        + n_l T_F 
        \biggl(
          \frac{64}{27}
          - \frac{8}{9} l_\mu
        \biggr)
      \Biggr]
        \nonumber \\ && \mbox{}
      + \frac{\mu^2}{\left({m^{\text{OS}}}\right)^2}
      \Biggl[
        C_A 
        \biggl(
          - \frac{91}{36}
          + \frac{\pi^2}{12}
          + \frac{11}{12} l_\mu
        \biggr)
        + n_l T_F
        \biggl(
          \frac{13}{18}
          - \frac{1}{3} l_\mu
        \biggr)
      \Biggr]
    \Biggr\}
    + \left( \frac{\alpha_s^{(n_l)}}{\pi} \right)^3 C_F
    \Biggl\{
      \frac{\mu}{{m^{\text{OS}}}}
      \Biggl[
        C_A^2
        \biggl(
          - \frac{130867}{1944}
    \nonumber \\ &&\mbox{}
          + \frac{511 \pi^2}{162}
          + \frac{19 \zeta_3}{2} 
          - \frac{\pi^4}{18}
          + \biggl(
            \frac{2518}{81}
            - \frac{22 \pi^2}{27}
          \biggr) l_\mu
          - \frac{121}{27} l_\mu^2
        \biggr)
        + C_A n_l T_F
        \biggl(
          \frac{19453}{486}
          - \frac{104 \pi^2}{81} 
          - 2 \zeta_3
        \nonumber \\ &&\mbox{}
          + \biggl(
            - \frac{1654}{81}
            + \frac{8\pi^2}{27}
          \biggr) l_\mu
          + \frac{88}{27} l_\mu^2
        \biggr)
        + C_F n_l T_F
        \biggl(
          \frac{11}{4}
          - \frac{4 \zeta_3}{3} 
          - \frac{2}{3} l_\mu
        \biggr)
        + n_l^2 T_F^2
        \biggl(
          - \frac{1292}{243}
          + \frac{8\pi^2}{81}
          + \frac{256}{81} l_\mu
          - \frac{16}{27} l_\mu^2
        \biggr)
      \Biggr]
      \nonumber \\ &&\mbox{}
      + \frac{\mu^2}{\left({m^{\text{OS}}}\right)^2}
      \Biggl[
        C_A^2
        \biggl(
          - \frac{96295}{5184}
          + \frac{445 \pi^2}{432} 
          + \frac{57 \zeta_3}{16} 
          - \frac{\pi^4}{48}
          + \biggl(
            \frac{2155}{216}
            - \frac{11 \pi^2}{36} 
          \biggr) l_\mu
          - \frac{121}{72} l_\mu^2
        \biggr)
        + C_A n_l T_F
        \biggl(
          \frac{13699}{1296}
          - \frac{23 \pi^2}{54}
      \nonumber \\ &&\mbox{}
          - \frac{3 \zeta_3}{4}
          + \biggl(
            - \frac{695}{108}
            + \frac{\pi^2}{9}
          \biggr) l_\mu
          + \frac{11}{9} l_\mu^2
        \biggr)
        + C_F n_l T_F
        \biggl(
          \frac{29}{32}
          - \frac{\zeta_3}{2}
          - \frac{1}{4} l_\mu
        \biggr)
        + n_l^2 T_F^2
        \biggl(
          - \frac{209}{162}
          + \frac{\pi^2}{27}
          + \frac{26}{27} l_\mu
          - \frac{2}{9} l_\mu^2
        \biggr)
      \Biggr]
    \Biggr\}\,,
    \label{eq::mKINomOS}
\end{eqnarray}
\end{widetext}
where $l_\mu = \ln\frac{2\mu}{\mu_s}$, $\mu$ denotes the Wilsonian cut-off and
$\mu_s$ is the renormalization scale of the strong coupling constant. The
colour factors of the SU$(N_C)$ gauge group are given by
$C_F=(N_C^2-1)/(2 N_C)$, $C_A=N_C$ and $T_F=1/2$ and the strong coupling
constant is defined in the $n_l$ flavour theory, where $n_l$ denotes the
number of light quark fields. Note that in our calculation no effects of
  finite charm quark masses are taken into account. The two-loop
  result of Eq.~(\ref{eq::mKINomOS}) and the $n_l^2$ term at three loops agree with
  Ref.~\cite{Czarnecki:1997sz}.

Next we replace the pole mass on the r.h.s. of Eq.~(\ref{eq::mKINomOS})
by the $\overline{\rm MS}$ mass using results up to three loops~\cite{Chetyrkin:1999ys,Chetyrkin:1999qi,Melnikov:2000qh}.
Also here we use $\alpha_s^{(n_l)}$ as the expansion parameter. In order to obtain
compact expressions we identify the renormalization scales of the
$\overline{\rm MS}$ parameters $\alpha_s$ and $\overline{m}$ and furthermore
specify the colour factors to QCD ($N_C=3$). This leads to
\begin{widetext}
\begin{eqnarray}
  \frac{m^{\text{kin}}}{\overline{m}} &=& 
  1
  + \frac{\alpha_s^{(n_l)}}{\pi}
  \biggl(
    \frac{4}{3}
    + l_m
    - \frac{16}{9} \frac{\mu}{\overline{m}}
    - \frac{2}{3} \frac{\mu^2}{\overline{m}^2}
  \biggr)
  + \left(\frac{\alpha_s^{(n_l)}}{\pi}\right)^2
  \Biggl\{
    \frac{307}{32}
    + \frac{\pi^2}{3}
    - \frac{\zeta_3}{6}
    + \frac{\pi^2}{9} l_2
    + \frac{509}{72} l_m
    + \frac{47}{24} l_m^2
    - n_l 
    \biggl(
      \frac{71}{144}
      + \frac{\pi^2}{18}
    \nonumber \\ && \mbox{}
      + \frac{13}{36} l_m
      + \frac{1}{12} l_m^2
    \biggr)
    + \frac{\mu}{\overline{m}}
    \biggl[
      - \frac{860}{27}
      + \frac{8\pi^2}{9}
      + \frac{88}{9} l_\mu
      + n_l
      \biggl(
        \frac{128}{81}
        - \frac{16}{27} l_\mu
      \biggr)
    \biggr]
    + \frac{\mu^2}{\overline{m}^2}
    \biggl[
      - \frac{83}{9}
      + \frac{\pi^2}{3}
      + \frac{2}{3} l_m
      + \frac{11}{3} l_\mu
  \nonumber \\ && \mbox{}
      + n_l
      \biggl(
        \frac{13}{27}
        - \frac{2}{9} l_\mu
      \biggr)
    \biggr]
  \Biggr\}
  + \left(\frac{\alpha_s^{(n_l)}}{\pi}\right)^3
  \Biggl\{
    \frac{8462917}{93312}
    + \frac{652841\pi^2}{38880}
    + \frac{58 \zeta_3}{27}
    - \frac{695 \pi^4}{7776} 
    - \frac{220 a_4}{27}
    - \frac{1439 \pi^2 \zeta_3}{432}
    + \frac{1975 \zeta_5}{216}
  \nonumber \\ && \mbox{}
    - \frac{575\pi^2}{162} l_2
    - \frac{22 \pi^2}{81} l_2^2
    - \frac{55}{162} l_2^4
    + l_m 
    \biggl(
      \frac{93391}{1296}
      + \frac{13 \pi^2}{6}
      - \frac{23 \zeta_3}{12}
      + \frac{13 \pi^2}{18} l_2
    \biggr)
    + \frac{21715}{864} l_m^2
    + \frac{1861}{432} l_m^3
    + n_l
    \biggl[
      - \frac{231847}{23328}
  \nonumber \\ && \mbox{}
      - \frac{991\pi^2}{648}
      - \frac{241 \zeta_3}{72}
      + \frac{61 \pi^4}{1944}
      + \frac{8 a_4}{27}
      - \frac{11\pi^2}{81} l_2 
      + \frac{2\pi^2}{81} l_2^2
      + \frac{1}{81} l_2^4
      - l_m
      \biggl(
        \frac{5171}{648}
        + \frac{17\pi^2}{36}
        + \frac{7 \zeta_3}{9}
        + \frac{\pi^2}{27} l_2
      \biggr)
      - \frac{385}{144} l_m^2
  \nonumber \\ && \mbox{}
      - \frac{43}{108} l_m^3
    \biggr]
    + n_l^2
    \biggl[
      \frac{2353}{23328}
      + \frac{13 \pi^2}{324}
      + \frac{7 \zeta_3}{54}
      + l_m 
      \biggl(
        \frac{89}{648}
        + \frac{\pi^2}{54}
      \biggr)
      + \frac{13}{216} l_m^2
      + \frac{1}{108} l_m^3
    \biggr]
    + \frac{\mu}{\overline{m}}
    \biggl[
      -\frac{130867}{162}
      +\frac{1022\pi^2}{27}
      \nonumber \\ && \mbox{}
      +114\zeta_3
      -\frac{2\pi^4}{3}
      + l_\mu
      \biggl(
        \frac{10072}{27}
        -\frac{88\pi^2}{9}
      \biggr)
      -\frac{484}{9} l_\mu^2
      + n_l
      \biggl(
        \frac{20047}{243}
        -\frac{208\pi^2}{81}
        -\frac{140\zeta_3}{27}
        + l_\mu
        \biggl(
          -\frac{3356}{81}
          +\frac{16\pi^2}{27}
        \biggr)
      \nonumber \\ &&\mbox{}
        +\frac{176}{27} l_\mu^2
      \biggr)
      + n_l^2
      \biggl(
        -\frac{1292}{729}
        +\frac{8\pi^2}{243}
        +\frac{256}{243} l_\mu
        -\frac{16}{81} l_\mu^2
      \biggr)
    \biggr]
    + \frac{\mu^2}{\overline{m}^2}
    \biggl[
      -\frac{22055}{108}
      +\frac{437\pi^2}{36}
      +\frac{1535\zeta_3}{36}
      -\frac{\pi^4}{4}
      +\frac{2\pi^2}{27} l_2
      \nonumber \\ && \mbox{}
      + l_m
      \biggl(
        \frac{1409}{108}
        -\frac{\pi^2}{3}
      \biggr)
      + l_\mu
      \biggl(
        \frac{689}{6}
        -\frac{11\pi^2}{3}
      \biggr)
      -\frac{11}{3} l_m l_\mu
      +\frac{23}{36} l_m^2
      -\frac{121}{6} l_\mu^2 
      + n_l
      \biggl(
        \frac{1699}{81}
        -\frac{8\pi^2}{9}
        -\frac{35\zeta_3}{18}
        -\frac{13}{18} l_m
        \nonumber \\ && \mbox{}
        + l_\mu
        \biggl(
          -\frac{691}{54}
          +\frac{2\pi^2}{9}
        \biggr)
        +\frac{2}{9} l_m l_\mu
        -\frac{1}{18} l_m^2
        + \frac{22}{9} l_\mu^2
      \biggr)
      + n_l^2
      \biggl(
        -\frac{209}{486}
        +\frac{\pi^2}{81}
        +\frac{26}{81} l_\mu
        -\frac{2}{27} l_\mu^2
      \biggr)
    \biggr]
  \Biggr\}\,,
    \label{eq::mKINomMS}
\end{eqnarray}
\end{widetext}
with $\overline{m}=\overline{m}(\mu_s)$ and
\begin{eqnarray}
  l_m = \ln\frac{\mu_s^2}{\overline{m}^2},\,
  l_2 = \ln2,\,
  a_4 = \text{Li}_4\left(\frac{1}{2}\right)\,.
\end{eqnarray}

We are now in the position to specify our results  to the charm and bottom
quark systems and check the perturbative stability of the quark mass relations.

The input values for our numerical analysis are
$\alpha_s^{(5)}(M_Z)=0.1179$~\cite{Tanabashi:2018oca},
$\overline{m}_c(3~\mbox{GeV})=0.993$~GeV~\cite{Chetyrkin:2017lif} and
$\overline{m}_b(\overline{m}_b)=4.163$~GeV~\cite{Chetyrkin:2009fv}.  We use {\tt RunDec}~\cite{Herren:2017osy}
for the running of the $\overline{\rm MS}$ parameters and the decoupling of
heavy particles. For the Wilsonian cut-off we choose $\mu=1$~GeV for bottom~\cite{Gambino:2013rza}
and $\mu=0.5$~GeV for charm~\cite{Gambino:2010jz}.

Let us start with the charm quark where we have $n_l=3$. We aim for a relation
between $m_c^{\rm kin}$ and $\overline{m}_c(\mu_s)$ for different choices of
$\mu_s$. Often numerical values for $\overline{m}_c(\overline{m}_c)$
are provided. However, this choice suffers from small renormalization scales
of the order 1~GeV. A more appropriate choice is thus $\overline{m}_c(2~\mbox{GeV})$
or $\overline{m}_c(3~\mbox{GeV})$. For the three choices we
obtain the following perturbative expansions 
\begin{eqnarray}
  m_c^{\rm kin} &=&  \hphantom{0}993 + 191 + 100 + 52~\mbox{MeV} = 1336~\mbox{MeV}\,,
                    \nonumber\\
  m_c^{\rm kin} &=&  1099 + 163 + \hphantom{0}76 + 34~\mbox{MeV} = 1372~\mbox{MeV}\,,
                    \nonumber\\
  m_c^{\rm kin} &=&  1279 + \hphantom{0}{84} + \hphantom{0}{30}
                    + 11~\mbox{MeV} = 1404~\mbox{MeV}\,,
\end{eqnarray} 
where from top to bottom $\mu_s=3~\mbox{GeV}, 2~\mbox{GeV}$ and
$\overline{m}_c$ have been chosen.  Within each equation the four numbers after
the first equality sign refer to the tree-level results and the one-, two- and
three-loop corrections.  One observes that for each choice of $\mu_s$ the
perturbative expansion behaves reasonably. The three-loop terms range from
$11$~MeV to $52$~MeV and roughly cover the splitting of the final numbers
for $m_c^{\rm kin}(0.5~\mbox{GeV})$.

In the case of the bottom quark we follow Ref.~\cite{Gambino:2011cq} and adapt two
different schemes for the charm quark: we either consider the charm quark  as
decoupled and set $n_l=3$, or we set $n_l=4$ which corresponds  to $m_c=0$.
(In the latter case one could include $m_c/m_b$ corrections which we postpone
to a future analysis~\cite{Fael_etal}.)

Using $\overline{m}_b(\overline{m}_b)$ as input we obtain the following results for the kinetic mass
\begin{eqnarray}
  m_b^{\rm kin} &=&  4163 + 248 + 81 + 30~\mbox{MeV} = 4521~\mbox{MeV}\,,
                    \nonumber\\
  m_b^{\rm kin} &=&  4163 + 259 + 77 + 25~\mbox{MeV} = 4523~\mbox{MeV}\,,
                    \label{eq::mbMS2KIN}
\end{eqnarray}
where the top and bottom line correspond to $n_l=3$ and $n_l=4$, respectively.
In both cases we observe a good convergence of the perturbative series: the
coefficients reduce by factors between $\approx2.5$ and $\approx3.5$ when
including higher orders. We suggest to estimate the unknown four-loop
corrections and contributions from higher dimensional operators, which scale
as $\alpha_s \mu^3/m_b^3 \sim \alpha_s^4$, by 50\% of the three-loop
corrections and assign an uncertainty of $15$~MeV and $12$~MeV for $n_l=3$ and
$n_l=4$, respectively.  Note, that our $\overline{m}_b$--$m_b^{\rm kin}$
scheme-conversion uncertainties are now smaller than the error of
$m_b^{\rm kin}$ as determined by global fits:
$m_b^{\rm kin}(1~\mbox{GeV}) = 4554 \pm 18$ MeV~\cite{Amhis:2019ckw}.

For the computation of $\overline{m}_b(\overline{m}_b)$ from the kinetic mass
we proceed as follows: we first use the inverted version of
Eq.~(\ref{eq::mKINomMS}) to compute the $\overline{\rm MS}$ bottom quark mass
at the scale $\mu_s=m_b^{\rm kin}$. Afterwards, we use the QCD renormalization
group equations at five-loop
accuracy~\cite{Baikov:2014qja,Luthe:2016xec,Baikov:2017ujl,Baikov:2016tgj,Herzog:2017ohr,Luthe:2017ttg,Chetyrkin:2017bjc}
as implemented in {\tt RunDec}~\cite{Herren:2017osy} to run to
$\mu_s=\overline{m}_b$.  In order to demonstrate the perturbative series we
choose $m_b^{\rm kin}$ from Eq.~(\ref{eq::mbMS2KIN}) and obtain 
for $n_l=3$ and $n_l=4$
\begin{eqnarray}
  \overline{m}_b(m_b^{\rm kin}) &=&  4521  - 273 - 101 - 39~\mbox{MeV}
\,,
                    \nonumber\\
  \overline{m}_b(m_b^{\rm kin}) &=&  4523 -  286 - \hphantom{0}98 - 34~\mbox{MeV}
\,,
\end{eqnarray}
with similar convergence properties as in Eq.~(\ref{eq::mbMS2KIN}).
Thus we estimate the uncertainty from unknown higher order
corrections as $\pm18$~MeV and $\pm17$~MeV, respectively.
In an alternative approach one can estimate the uncertainty from
the variation of the intermediate scale $\mu_s$ which
leads to similar uncertainty estimates.

Finally, we present simple formulae which can be used to convert the
scale-invariant bottom quark mass to the kinetic scheme or vice versa
using the preferred input values for the mass and strong coupling constant.
We have
\begin{eqnarray}
  \frac{\overline{m}_b(\overline{m}_b)}{\mbox{MeV}} &=& 4163 +
                    \Delta^{(n_l)}_{\rm kin} \{13,13\} - \Delta_{\alpha_s} \{7,7\} 
                   \pm \{18,17\}
                                     \,, 
                                     \nonumber\\
  \frac{m_b^{\rm kin}}{\mbox{MeV}} &=& 4522 
                    + \Delta_{\overline{\rm MS}} \{18,18\} +
                    \Delta_{\alpha_s} \{8,8\}
                   \pm \{15,12\}
                    \,,
                    \nonumber\\
\end{eqnarray}
where the first (second) number in the curly brackets corresponds to $n_l=3$
($n_l=4$). Furthermore, we have defined
$\Delta^{(3)}_{\rm kin} = (m_b^{\rm kin}/\mbox{MeV}-4518)/15$, 
$\Delta^{(4)}_{\rm kin} = (m_b^{\rm kin}/\mbox{MeV}-4520)/15$, 
$\Delta_{\overline{\rm MS}}= (\overline{m}_b(\overline{m}_b)/\mbox{MeV}-4163)/16$ and 
$\Delta_{\alpha_s}= (\alpha_s-0.1179)/0.001$.


\bigskip
{\bf Conclusions.}
The main purpose of this Letter is the improvement of the precision in the
conversion relation between the heavy quark kinetic and $\overline{\rm MS}$
masses. This goal is reached by computing the relation between the kinetic and
pole mass to three-loop order; previously only two-loop corrections, supplemented
by large-$\beta_0$ terms, were available.  The main results of this paper
can be found in Eqs.~(\ref{eq::mKINomOS}) and~(\ref{eq::mKINomMS}).  Using a
conservative uncertainty estimate the new corrections reduces the uncertainty
in transformation formulas by about a factor two.  Our findings constitute
important ingredients in the extraction of $|V_{cb}|$ at the percent level or
even below.


\smallskip

{\bf Acknowledgements.}  We thank Andrzej Czarnecki and Miko{\l}aj Misiak for
useful discussions and communications and Alexander Smirnov for help in the
use of {\tt asy}~\cite{Pak:2010pt}. 
We also thank Joshua Davies and Thomas Mannel for carefully reading the
manuscript.
We are grateful to Florian Herren for
providing us his program which automates the partial fraction decomposition in
case of linearly dependent denominators.  This research was supported by the
Deutsche Forschungsgemeinschaft (DFG, German Research Foundation) under grant
396021762 --- TRR 257 ``Particle Physics Phenomenology after the Higgs
Discovery''.


\begin{thebibliography}{99}

%
%

\bibitem{Amhis:2019ckw}
Y.~S.~Amhis \textit{et al.} [HFLAV],
[arXiv:1909.12524 [hep-ex]].

\bibitem{Buras:1997fb}
A.~J.~Buras and R.~Fleischer,
Adv. Ser. Direct. High Energy Phys. \textbf{15} (1998), 65-238
%
[arXiv:hep-ph/9704376 [hep-ph]].

\bibitem{Brod:2010hi}
J.~Brod, M.~Gorbahn and E.~Stamou,
Phys. Rev. D \textbf{83} (2011), 034030
%
[arXiv:1009.0947 [hep-ph]].

\bibitem{Bobeth:2013uxa}
C.~Bobeth, M.~Gorbahn, T.~Hermann, M.~Misiak, E.~Stamou and M.~Steinhauser,
Phys. Rev. Lett. \textbf{112} (2014), 101801
%
[arXiv:1311.0903 [hep-ph]].

\bibitem{Ligeti:2016qpi}
Z.~Ligeti and F.~Sala,
JHEP \textbf{09} (2016), 083
%
[arXiv:1602.08494 [hep-ph]].

\bibitem{Gambino:2013rza}
  P.~Gambino and C.~Schwanda,
  Phys.\ Rev.\ D {\bf 89} (2014) no.1,  014022
%
  [arXiv:1307.4551 [hep-ph]].

\bibitem{Alberti:2014yda}
A.~Alberti, P.~Gambino, K.~J.~Healey and S.~Nandi,
Phys. Rev. Lett. \textbf{114} (2015) no.6, 061802
%
[arXiv:1411.6560 [hep-ph]].

\bibitem{Gambino:2016jkc}
P.~Gambino, K.~J.~Healey and S.~Turczyk,
Phys. Lett. B \textbf{763} (2016), 60-65
%
[arXiv:1606.06174 [hep-ph]].

\bibitem{Bigi:1996si}
I.~I.~Bigi, M.~A.~Shifman, N.~Uraltsev and A.~I.~Vainshtein,
Phys. Rev. D \textbf{56} (1997), 4017-4030
%
[arXiv:hep-ph/9704245 [hep-ph]].

\bibitem{Hoang:1999zc}
A.~Hoang and T.~Teubner,
Phys. Rev. D \textbf{60} (1999), 114027
%
[arXiv:hep-ph/9904468 [hep-ph]].

\bibitem{Hoang:1998hm}
A.~H.~Hoang, Z.~Ligeti and A.~V.~Manohar,
Phys. Rev. D \textbf{59} (1999), 074017
%
[arXiv:hep-ph/9811239 [hep-ph]].

\bibitem{Hoang:1998ng}
A.~H.~Hoang, Z.~Ligeti and A.~V.~Manohar,
Phys. Rev. Lett. \textbf{82} (1999), 277-280
%
[arXiv:hep-ph/9809423 [hep-ph]].

\bibitem{Bauer:2004ve}
C.~W.~Bauer, Z.~Ligeti, M.~Luke, A.~V.~Manohar and M.~Trott,
Phys. Rev. D \textbf{70} (2004), 094017
%
[arXiv:hep-ph/0408002 [hep-ph]].

\bibitem{Marquard:2015qpa}
  P.~Marquard, A.~V.~Smirnov, V.~A.~Smirnov and M.~Steinhauser,
  Phys.\ Rev.\ Lett.\  {\bf 114} (2015) no.14,  142002
%
  [arXiv:1502.01030 [hep-ph]].

\bibitem{Marquard:2016dcn}
  P.~Marquard, A.~V.~Smirnov, V.~A.~Smirnov, M.~Steinhauser and D.~Wellmann,
  Phys.\ Rev.\ D {\bf 94} (2016) no.7,  074025
%
  [arXiv:1606.06754 [hep-ph]].

\bibitem{Czarnecki:1997sz}
  A.~Czarnecki, K.~Melnikov and N.~Uraltsev,
  Phys.\ Rev.\ Lett.\  {\bf 80} (1998) 3189
%
  [hep-ph/9708372].

\bibitem{Beneke:1994sw}
  M.~Beneke and V.~M.~Braun,
  Nucl.\ Phys.\ B {\bf 426} (1994) 301
%
  [hep-ph/9402364].

\bibitem{Bigi:1994em}
  I.~I.~Y.~Bigi, M.~A.~Shifman, N.~G.~Uraltsev and A.~I.~Vainshtein,
  Phys.\ Rev.\ D {\bf 50} (1994) 2234
%
  [hep-ph/9402360].

\bibitem{Melnikov:2000qh}
  K.~Melnikov and T.~v.~Ritbergen,
  Phys.\ Lett.\ B {\bf 482} (2000) 99
%
  [hep-ph/9912391].

\bibitem{Bigi:1994ga}
  I.~I.~Y.~Bigi, M.~A.~Shifman, N.~G.~Uraltsev and A.~I.~Vainshtein,
  Phys.\ Rev.\ D {\bf 52} (1995) 196
%
  [hep-ph/9405410].

\bibitem{Gambino:2011cq}
P.~Gambino,
JHEP \textbf{09} (2011), 055
%
[arXiv:1107.3100 [hep-ph]].

\bibitem{Fael_etal}
M.~Fael, K.~Sch\"onwald and M.~Steinhauser,
[arXiv:2011.11655 [hep-ph]].

\bibitem{Beneke:1997zp}
  M.~Beneke and V.~A.~Smirnov,
  Nucl.\ Phys.\ B {\bf 522} (1998) 321
%
  [hep-ph/9711391].

\bibitem{Smirnov:2012gma}
  V.~A.~Smirnov,
  Springer Tracts Mod.\ Phys.\  {\bf 250} (2012) 1.
%

\bibitem{Nogueira:1991ex}
  P.~Nogueira,
  J.\ Comput.\ Phys.\  {\bf 105} (1993) 279.
%

\bibitem{Ruijl:2017dtg}
B.~Ruijl, T.~Ueda and J.~Vermaseren,
[arXiv:1707.06453 [hep-ph]].

\bibitem{Pak:2010pt}
A.~Pak and A.~Smirnov,
Eur.\ Phys.\ J.\ C \textbf{71} (2011), 1626
%
[arXiv:1011.4863 [hep-ph]].

\bibitem{Smirnov:2019qkx}
  A.~V.~Smirnov and F.~S.~Chuharev,
%
  arXiv:1901.07808 [hep-ph].

\bibitem{Lee:2012cn}
  R.~N.~Lee,
  arXiv:1212.2685 [hep-ph];
  R.~N.~Lee,
  J.\ Phys.\ Conf.\ Ser.\  {\bf 523} (2014) 012059
%
  [arXiv:1310.1145 [hep-ph]].

\bibitem{Schneider:2007}
C.~Schneider, {S\'em.~Lothar. Combin.\/} {\bf 56} (2007) 1,
 article B56b;
C.~Schneider, in:~{{Computer Algebra in Quantum Field Theory: Integration,
  Summation and Special Functions}\/} Texts and Monographs in Symbolic
  Computation eds. C.~Schneider and J.~Bl\"umlein  (Springer, Wien, 2013) 325
  arXiv:1304.4134 [cs.SC].

\bibitem{Ablinger:2010pb}
  J.~Ablinger, J.~Bl\"umlein, S.~Klein and C.~Schneider,
  Nucl.\ Phys.\ Proc.\ Suppl.\  {\bf 205-206} (2010) 110
  [arXiv:1006.4797 [math-ph]];
  J.~Bl\"umlein, A.~Hasselhuhn and C.~Schneider,
  PoS (RADCOR 2011) 032
  [arXiv:1202.4303 [math-ph]];
  C.~Schneider,
  J.\ Phys.\ Conf.\ Ser.\  {\bf 523} (2014) 012037
  [arXiv:1310.0160 [cs.SC]].

\bibitem{HarmonicSums}
J.~Vermaseren,
Int. J. Mod. Phys. A \textbf{14} (1999), 2037-2076
%
[arXiv:hep-ph/9806280 [hep-ph]];
E.~Remiddi and J.~Vermaseren,
Int. J. Mod. Phys. A \textbf{15} (2000), 725-754
%
[arXiv:hep-ph/9905237 [hep-ph]];
J.~Bl\"umlein,
Comput. Phys. Commun. \textbf{180} (2009), 2218-2249
%
[arXiv:0901.3106 [hep-ph]];
  J.~Ablinger,
  Diploma Thesis, J. Kepler University Linz, 2009,
  arXiv:1011.1176 [math-ph];
  J.~Ablinger, J.~Bl\"umlein and C.~Schneider,
  J.\ Math.\ Phys.\  {\bf 52} (2011) 102301
  [arXiv:1105.6063 [math-ph]];
J.~Ablinger, J.~Bl\"umlein and C.~Schneider,
J. Math. Phys. \textbf{54} (2013), 082301
%
[arXiv:1302.0378 [math-ph]];
  J.~Ablinger,
  Ph.D. Thesis, J. Kepler University Linz, 2012,
  arXiv:1305.0687 [math-ph];
J.~Ablinger, J.~Bl\"umlein and C.~Schneider,
J. Phys. Conf. Ser. \textbf{523} (2014), 012060
%
[arXiv:1310.5645 [math-ph]];
J.~Ablinger, J.~Bl\"umlein, C.~Raab and C.~Schneider,
J. Math. Phys. \textbf{55} (2014), 112301
%
[arXiv:1407.1822 [hep-th]];
%
J.~Ablinger,
PoS \textbf{LL2014} (2014), 019
%
[arXiv:1407.6180 [cs.SC]];
%
J.~Ablinger,
[arXiv:1606.02845 [cs.SC]];
%
J.~Ablinger,
PoS \textbf{RADCOR2017} (2017), 069
[arXiv:1801.01039 [cs.SC]];
%
J.~Ablinger,
PoS \textbf{LL2018} (2018), 063;
%
%
J.~Ablinger,
[arXiv:1902.11001 [math.CO]].
%

\bibitem{PSLQ}
H.R.P.~Ferguson and D.H.~Bailey, RNR Technical Report, RNR-91-032;
H.R.P.~Ferguson, D.H.~Bailey and S.~Arno, NASA Technical Report,
NAS-96-005.

\bibitem{Kotikov:1990kg}
A.~Kotikov,
Phys. Lett. B \textbf{254} (1991), 158-164

\bibitem{Gehrmann:1999as}
T.~Gehrmann and E.~Remiddi,
Nucl. Phys. B \textbf{580} (2000), 485-518
[arXiv:hep-ph/9912329 [hep-ph]].

\bibitem{Henn:2013pwa}
J.~M.~Henn,
Phys. Rev. Lett. \textbf{110} (2013), 251601
[arXiv:1304.1806 [hep-th]].

\bibitem{Chetyrkin:1999ys}
  K.~G.~Chetyrkin and M.~Steinhauser,
  Phys.\ Rev.\ Lett.\  {\bf 83} (1999) 4001
%
  [hep-ph/9907509].

\bibitem{Chetyrkin:1999qi}
  K.~G.~Chetyrkin and M.~Steinhauser,
  Nucl.\ Phys.\ B {\bf 573} (2000) 617
%
  [hep-ph/9911434].

\bibitem{Tanabashi:2018oca}
M.~Tanabashi \textit{et al.} [Particle Data Group],
Phys. Rev. D \textbf{98} (2018) no.3, 030001
%

\bibitem{Chetyrkin:2017lif}
K.~G.~Chetyrkin, J.~H.~K\"uhn, A.~Maier, P.~Maierhofer, P.~Marquard,
M.~Steinhauser and C.~Sturm,
%
[arXiv:1710.04249 [hep-ph]].

\bibitem{Chetyrkin:2009fv}
K.~Chetyrkin, J.~K\"uhn, A.~Maier, P.~Maierhofer, P.~Marquard, M.~Steinhauser
and C.~Sturm,
Phys. Rev. D \textbf{80} (2009), 074010
%
[arXiv:0907.2110 [hep-ph]].

\bibitem{Herren:2017osy}
F.~Herren and M.~Steinhauser,
Comput. Phys. Commun. \textbf{224} (2018), 333-345
%
[arXiv:1703.03751 [hep-ph]].

\bibitem{Gambino:2010jz}
P.~Gambino and J.~F.~Kamenik,
Nucl. Phys. B \textbf{840} (2010), 424-437
%
[arXiv:1004.0114 [hep-ph]].

\bibitem{Baikov:2014qja}
P.~Baikov, K.~Chetyrkin and J.~K\"uhn,
JHEP \textbf{10} (2014), 076
%
[arXiv:1402.6611 [hep-ph]].

\bibitem{Luthe:2016xec}
T.~Luthe, A.~Maier, P.~Marquard and Y.~Schroder,
JHEP \textbf{01} (2017), 081
%
[arXiv:1612.05512 [hep-ph]].

\bibitem{Baikov:2017ujl}
P.~Baikov, K.~Chetyrkin and J.~K\"uhn,
JHEP \textbf{04} (2017), 119
%
[arXiv:1702.01458 [hep-ph]].

\bibitem{Baikov:2016tgj}
  P.~A.~Baikov, K.~G.~Chetyrkin and J.~H.~K\"uhn,
  Phys.\ Rev.\ Lett.\  {\bf 118} (2017) no.8,  082002
%
  [arXiv:1606.08659 [hep-ph]].

\bibitem{Herzog:2017ohr}
  F.~Herzog, B.~Ruijl, T.~Ueda, J.~A.~M.~Vermaseren and A.~Vogt,
  JHEP {\bf 1702} (2017) 090
%
  [arXiv:1701.01404 [hep-ph]].

\bibitem{Luthe:2017ttg}
  T.~Luthe, A.~Maier, P.~Marquard and Y.~Schroder,
  JHEP {\bf 1710} (2017) 166
%
  [arXiv:1709.07718 [hep-ph]].

\bibitem{Chetyrkin:2017bjc}
K.~Chetyrkin, G.~Falcioni, F.~Herzog and J.~Vermaseren,
JHEP \textbf{10} (2017), 179
%
[arXiv:1709.08541 [hep-ph]].


\end{thebibliography}
\end{document}